\documentclass[aps, prd, superscriptaddress, preprintnumbers, floatfix]{revtex4}

\usepackage{graphicx,amsfonts,amsmath,amssymb,amstext}
\usepackage{float,wrapfig}
\usepackage{subfigure, psfrag}
\usepackage{dsfont}
\usepackage{color}
\usepackage{hyperref}
\hypersetup{
	colorlinks = true,
	linkcolor = {blue},
	citecolor = {blue},
    urlcolor = {blue}
}
\usepackage{bm}

\newcommand{\be}{\begin{equation}}
\newcommand{\ee}{\end{equation}}
\newcommand{\ba}{\begin{eqnarray}}
\newcommand{\ea}{\end{eqnarray}}
\newcommand{\sign}[1]{\,\mbox{sgn}\left({#1}\right)}

\definecolor{purple}{rgb}{0.8,0,0.6}
\definecolor{darkgreen}{rgb}{0.00,0.6,0.00}

\extrafloats{100}

\begin{document}

\title{Superconductivity in Weyl semimetals in a strong pseudomagnetic field}
\date{July 27, 2020}

\author{P. O. Sukhachov}
\affiliation{Nordita, KTH Royal Institute of Technology and Stockholm University, Roslagstullsbacken 23, SE-106 91 Stockholm, Sweden}

\author{E. V. Gorbar}
\affiliation{Department of Physics, Taras Shevchenko National Kyiv University, Kyiv, 03680, Ukraine}
\affiliation{Bogolyubov Institute for Theoretical Physics, Kyiv, 03680, Ukraine}

\begin{abstract}
The superconducting $s$-wave state in Weyl semimetals in a strong strain-induced pseudomagnetic field is investigated in a model with local four-fermion interaction. It is found that only the inter-node pairing is possible in the lowest pseudo-Landau level approximation. Unlike the case of the lowest Landau level in a conventional magnetic field, the corresponding gap equation has only a trivial solution.
Nevertheless, superconductivity can be induced via the proximity effect with a usual $s$-wave spin-singlet superconductor. Since a pseudomagnetic field is present necessarily at the surface of a Weyl semimetal, the proximity effect is strongly affected by the pseudomagnetic field. The analysis of such an effect showed that while no gap is opened in the spectrum, the degeneracy of energy levels is lifted. The unique character of the proximity effect in Weyl semimetals can be probed via the density of states, the spectral function, and the tunneling current. The density of states does not vanish at small energies and scales linearly with the pseudomagnetic field strength. This scaling is manifested also in the tunneling current.
\end{abstract}

\maketitle

\section{Introduction}
\label{sec:introduction}

The interplay of superconductivity and magnetic fields is a nontrivial problem with a rich history as well as high fundamental and applied impact. The relevant phenomena are the Meissner effect connected with the expulsion of a magnetic field from a superconductor~\cite{Meissner-Ochsenfeld:1933}, the Abrikosov vortex state in type-II superconductors~\cite{Abrikosov:1957}, the spatially nonuniform Larkin--Ovchinnikov--Fulde--Ferrell (LOFF) phase~\cite{LOFF:1964}, etc. Although they are quite diverse, all these effects show that magnetic field suppresses superconductivity. Indeed, a strong enough magnetic field destroys the superconducting state because the kinetic energy of diamagnetic currents, which expel the external magnetic field, becomes too large. Therefore, it is rather surprising that superconductivity was suggested to revive in a strong magnetic field where the Landau levels form~\cite{Rasolt:1987,Dukan-Tesanovic:1991,Tesanovic:1992,Rosenstein:2017}.

The basic idea considered in Refs.~\cite{Rasolt:1987,Dukan-Tesanovic:1991,Tesanovic:1992,Rosenstein:2017} relies on the fact that the role of interactions increases as the energy dispersion grows slower with momentum. It is well known that magnetic field quenches the kinetic energy and effectively reduces the spatial dimension of a system by two. For example, the kinetic energy is quenched completely in two-dimensional (2D) systems where the Landau levels are flat. The systems with quenched kinetic energy present a perfect platform for realizing interaction-induced states. The fractional Hall effect~\cite{Tsui:1982} and the magnetic catalysis~\cite{Gusynin:1994,Miransky-Shovkovy:rev-2015} are two renowned examples. It is worth noting, however, that the back-reaction of superconducting currents on the magnetic field was not taken into account in Refs.~\cite{Rasolt:1987,Dukan-Tesanovic:1991,Tesanovic:1992,Rosenstein:2017}. Therefore, one should be careful in drawing rigorous conclusions and making experimental predictions. Indeed, until now, there are no definitive experimental evidences for superconductivity in a strong magnetic field.

The systems with relativisticlike dispersion relations are particularly advantageous for observing the unconventional superconductivity in strong magnetic fields~\cite{Rosenstein:2017} because the ultraquantum regime is routinely achievable there. A paradigmatic example of such three-dimensional (3D) relativisticlike systems are Dirac and Weyl semimetals~\cite{Yan-Felser:2017-Rev,Hasan-Huang:rev-2017,Armitage-Vishwanath:2017-Rev}. These semimetals are characterized by the band structure where the valence and conduction bands touch at isolated points in the Brillouin zone known as Dirac points and Weyl nodes, respectively. The corresponding low-energy quasiparticles are chiral fermions that have a linear dispersion relation and are described by relativisticlike Dirac and Weyl equations. According to the Nielsen--Ninomiya theorem~\cite{Nielsen-Ninomiya-1,Nielsen-Ninomiya-2}, Weyl nodes in solids always occur in pairs of opposite chiralities. The nodes are separated by $2\mathbf{b}$ in momentum space (the corresponding parameter is known as the chiral shift) and/or by $2b_0$ in energy. While $\mathbf{b}$ breaks the time-reversal (TR) symmetry, $b_0$ violates the parity inversion. Recently, by using ab initio calculations~\cite{Wang-Canfield:2019,Soh-Boothroyd:2019,Ma-Shi:2019} and angle-resolved photoemission spectroscopy (ARPES) measurements~\cite{Soh-Boothroyd:2019,Ma-Shi:2019}, it was shown that EuCd$_2$As$_2$ realizes a Weyl semimetal with broken TR symmetry. It features a single pair of Weyl nodes in the Brillouin zone near the Fermi level when alloyed with barium or an external magnetic field is applied.

Nontrivial topology of Weyl semimetals, whose Weyl nodes are sources and sinks of the Berry flux, plays an important role in their superconducting properties~\cite{Schnyder:rev-2015,Sato:2017}. In general, there exist two distinctive types of superconducting pairing that could be realized in Weyl semimetals~\cite{Cho-Moore:2012,Wei:2014vsa,Hosur:2014fba,Bednik:2015tha,Kobayashi:2015,Hashimoto-Sato:2016,Kim-Gilbert:2016}. The first is the \emph{inter-node} pairing of quasiparticles from the Weyl nodes of opposite chirality. The other type of pairing is the \emph{intra-node} one, which involves quasiparticles from the same Weyl node of a given chirality. The intra-node pairing leads to spin-singlet Cooper pairs with nonzero total momenta and, consequently, produces a LOFF-type state. On the other hand, the inter-node pairing might allow for spin-singlet and spin-triplet pairings with zero total momentum. The competition of intra- and inter-node pairing states in Weyl semimetals is subtle and depends on the model details. For example, it was argued that the inter-node pairing could be more favorable energetically than the intra-node one~\cite{Bednik:2015tha}, even though the former has point nodes in the gap function~\cite{Meng-Balents:2012,Cho-Moore:2012}.

Another important manifestation of the nontrivial topology of Weyl semimetals is the topologically protected surface Fermi arc states, which connect the projections of the bulk Weyl nodes of opposite chiralities onto the surface~\cite{Wan:2011udc}. Similarly to the case of topological insulators~\cite{Fu-Kane:2008}, these surface states modify the superconducting proximity effect. For example, it was shown~\cite{Chen:2016} that the Fermi arcs allows for a $p$-wave interface superconducting state with a single gapless Majorana mode in a Weyl semimetal with broken TR but intact mirror symmetry. Furthermore, it was shown that an exotic surface state with crossed flat bands in the superconducting state of a Weyl semimetal is possible due to the Fermi arcs~\cite{Lu-Tanaka:2016}.

In this study, we investigate a different possibility to realize an unconventional superconducting state in Weyl semimetals. It relies on strong pseudomagnetic fields generated by mechanical strains~\cite{Zhou-Shi:2013,Zubkov:2015,Cortijo-Vozmediano:2015} (see also Ref.~\cite{Ilan-Pikulin:rev-2019} for a review). Unlike their magnetic counterparts, these strain-induced fields interact with left- and right-handed fermions in Weyl semimetals with different sign. Moreover, pseudomagnetic fields appear at the surface of Weyl semimetals where the chiral shift abruptly changes even in the absence of a mechanical strain~\cite{Chernodub-Vozmediano:2014,Grushin:2016,Grushin:2018} leading to an effective axial gauge field. It was argued also that the Fermi arc surface states could be interpreted as lowest pseudo-Landau levels~\cite{Grushin:2016}. Since pseudoelectromagnetic fields have a completely different physical origin compared to usual electromagnetic fields, they do not induce diamagnetic currents that can back-react and destroy the superconducting state. This suggests that the Meissner effect should be absent for pseudomagnetic fields and implies that these fields may naively enhance and promote the superconductivity in Weyl semimetals.

It was found, however, that a weak pseudomagnetic field does not favor inter-node superconducting pairing in Weyl semimetals~\cite{Matsushita-Fujimoto:2018,Gorbar:2018pit}. Since only the regime of a weak pseudomagnetic field in the quasiclassical Eilenberger formalism was considered in Refs.~\cite{Matsushita-Fujimoto:2018,Gorbar:2018pit}, the question about the fate of superconductivity in the ultraquantum regime in a strong pseudomagnetic field remains open and provides one of the main motivations for the present study. While we found that a strong pseudomagnetic field does not support an intrinsic $s$-wave superconducting state, the proximity effect with a usual $s$-wave superconductor can still induce the superconductivity in Weyl semimetals. The corresponding superconducting state is unusual because, unlike the case of conventional superconductors, the density of states (DOS) does not vanish at small energies. This is explained by the fact that the inter-node $s$-wave pairing does not open a gap in the energy spectrum. Nevertheless, this pairing leads to a splitting of energy levels, which is evident from the spectral function. The effect of a pseudomagnetic field is imprinted in the linear in pseudomagnetic field dependence of the DOS and the tunneling current. As we will discuss below, the key difference between the superconducting pairing in strong pseudomagnetic and magnetic fields lies in a different structure of the lowest pseudo-Landau and conventional Landau levels, respectively, as well as in the role of the corresponding gaps.

This paper is organized as follows. Section~\ref{sec:pairing} is devoted to the study of intrinsic superconductivity in Weyl semimetals in a strong pseudomagnetic field. In particular, the Hamiltonian, wave functions, superconducting pairings, and the gap equation are defined. The proximity effect with a usual superconductor is considered in Sec.~\ref{sec:proximity}. As an experimentally accessible signatures, the spectral function, the DOS, and the tunneling current are discussed in Sec.~\ref{sec:tunneling-current}. Results are summarized in Sec.~\ref{sec:Summary}. The superconducting pairing in a strong magnetic field is considered in Appendix~\ref{sec:App-B}. A few technical details are presented in Appendix~\ref{sec:App-1}. Throughout this study, we set $\hbar=1$.

\section{Intrinsic superconductivity in strong pseudomagnetic field}
\label{sec:pairing}

In this section, the intrinsic superconductivity in a simple low-energy model of a Weyl semimetal subject to a strong pseudomagnetic field is investigated. We start by defining the Hamiltonian and wave functions. Further, the gap equation in a model with local four-fermion interaction is derived and solved.

\subsection{Model}
\label{sec:pairing-model}

Performing the Hubbard--Stratonovich transformation~\cite{Hubbard:1959,Stratonovich:1957}, the effective action for a model with local four-fermion interaction attains the following form:
\begin{equation}
S=\int dtd^3r \,\left[-i\mbox{Tr}\,\mbox{Ln}\,\left(i\partial_t - H_{\rm BdG}\right) +i\mbox{Tr}\,\mbox{Ln}\,\left(i\partial_t - H_{\rm BdG}\right)_{\Delta\to0, \mu\to0} - \frac{\mbox{tr}[\hat{\Delta}^{\dagger}\hat{\Delta}]}{g}\right],
\label{effective-action}
\end{equation}
where $g$ is the dimensionful coupling constant, $\hat{\Delta}$ is the superconducting gap matrix,
\begin{equation}
H_{\rm BdG}(\mathbf{k}) = \left(
                    \begin{array}{cc}
                      \hat{H}(\mathbf{k}) & \hat{\Delta} \\
                      \hat{\Delta}^{\dag} & -\hat{\Theta} \hat{H}(\mathbf{k})\hat{\Theta}^{-1} \\
                    \end{array}
                  \right)
\label{BdG-Hamiltonian}
\end{equation}
is the Bogolyubov--de Gennes (BdG) Hamiltonian, $\mathbf{k}$ is the momentum, and $\hat{\Theta}$ is the time-reversal operator. In addition, the trace and the logarithm in Eq.~(\ref{effective-action}) are taken in the functional sense.

\begin{figure*}[!ht]
\centering
\subfigure[]{\includegraphics[width=0.35\textwidth]{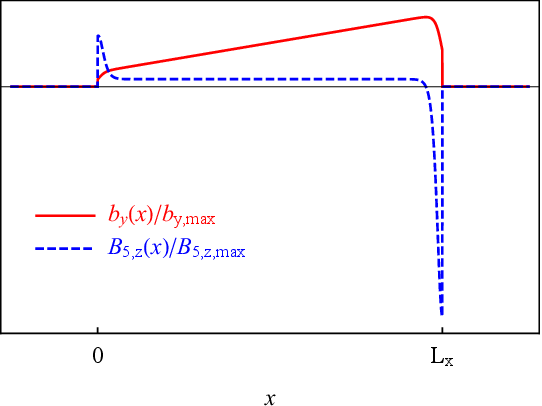}}
\hspace{0.1\textwidth}
\subfigure[]{\raisebox{0.075\textwidth}{\includegraphics[width=0.35\textwidth]{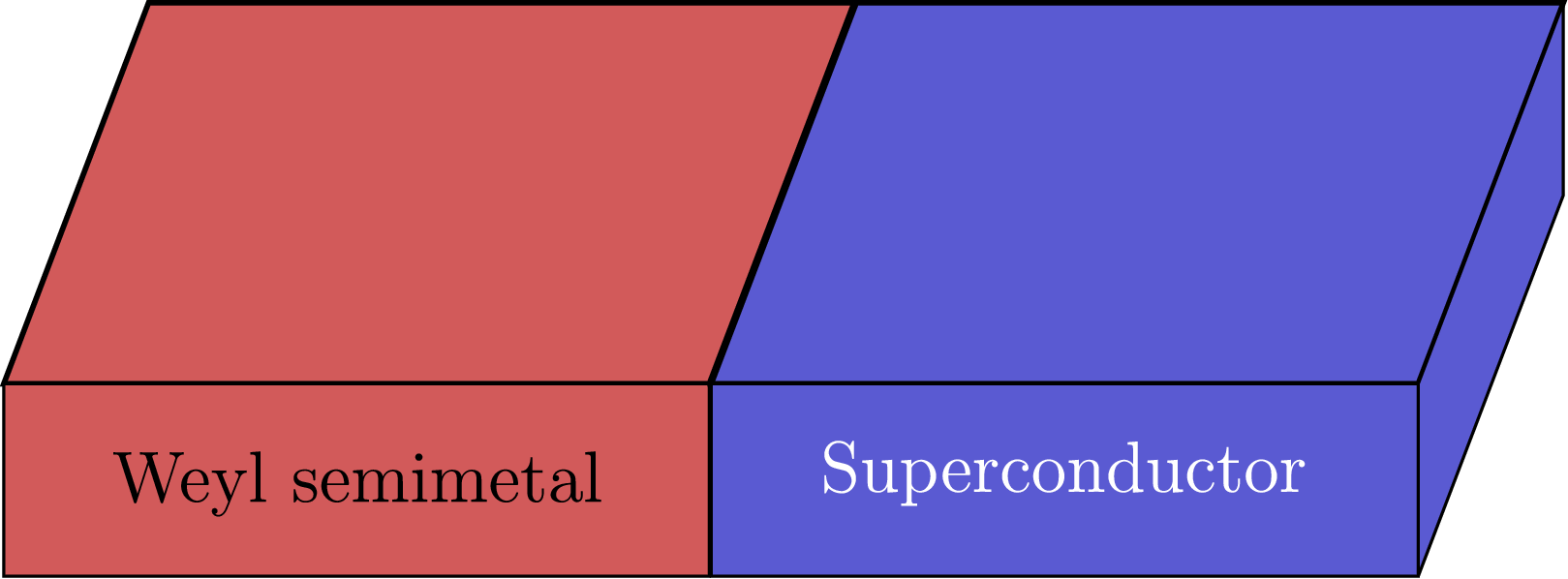}}}
\caption{Schematics of the chiral shift profile and a pseudomagnetic field in a slab of Weyl semimetal (panel (a)). Illustration of the proximity junction between a Weyl semimetal and a conventional superconducting metal (panel (b)).
}
\label{fig:model}
\end{figure*}

We consider the minimal model of Weyl semimetal with two Weyl nodes of opposite chiralities separated by $2\mathbf{b}$ in momentum space. The linearized Hamiltonian reads
\begin{equation}
\label{9-Model-hatH}
\hat{H} = \left(
                    \begin{array}{cc}
                      H_{+}  & 0 \\
                      0 & H_{-} \\
                    \end{array}
                  \right),
\end{equation}
where
\begin{equation}
\label{9-Model-H-chi}
H_{\lambda}=-\mu +\lambda v_F \bm{\sigma} \cdot\left(-i\bm{\nabla} + \lambda \frac{e}{c} \mathbf{A}_5 -\lambda \mathbf{b}\right).
\end{equation}
Here $\lambda=\pm$ is the chirality of Weyl nodes, $\mu$ is the electric chemical potential, $v_F$ is the Fermi velocity, $\bm{\sigma}$ is the vector of the Pauli matrices, $c$ is the speed of light, and $\mathbf{A}_5$ is the axial gauge field. This field can be induced by strains~\cite{Zhou-Shi:2013,Zubkov:2015,Cortijo-Vozmediano:2015}. Moreover, a coordinate-dependent axial gauge field appears necessarily at the surface of a Weyl semimetal, where the chiral shift terminates~\cite{Chernodub-Vozmediano:2014,Grushin:2016,Grushin:2018,Benito-Matias-Gonzalez:2020} [see also Fig.~\ref{fig:model}(a)]. This coordinate-dependent axial gauge field $\mathbf{A_5}$ gives rise to a pseudomagnetic field $\mathbf{B}_5=\bm{\nabla}\times\mathbf{A_5}$.

A schematic illustration of the chiral shift profile and the corresponding pseudomagnetic field is shown in Fig.~\ref{fig:model}(a). While, in general, the pseudomagnetic field is nonuniform, we focus in this section on a constant pseudomagnetic field in the bulk directed along the $z$ and defined by $\mathbf{A}_5=B_5 x\hat{\mathbf{y}}$, where $\hat{\mathbf{y}}$ is the unit vector along the $y$ direction. Furthermore, as will be discussed in Sec.~\ref{sec:proximity}, such a model is also relevant for investigating the proximity effect in a junction of a Weyl semimetal and a superconductor [see Fig.~\ref{fig:model}(b)].

Finally, for Hamiltonian (\ref{9-Model-hatH}), the TR operator has the following form:
\begin{equation}
\label{9-Model-Theta-def}
\hat{\Theta} = i \mathds{1}_2\otimes\sigma_y \hat{K} \Pi_{\mathbf{k}\to-\mathbf{k}}.
\end{equation}

To calculate the functional trace in the first term of the effective action~(\ref{effective-action}), let us to determine eigenstates and eigenenergies of the BdG Hamiltonian, which satisfy the following equation:
\begin{equation}
H_{\rm BdG}\Psi_{\rm BdG}=\epsilon\Psi_{\rm BdG}.
\label{BdG-eigenstate}
\end{equation}
The wave function $\Psi_{\rm BdG}$ consists of two parts corresponding to the wave function of the Weyl Hamiltonian and its TR copy. The first part reads
\begin{equation}
\label{9-Model-Nambu-Psi-1}
\Psi = \left\{\psi^+_{\uparrow} (\mathbf{k}),\psi^+_{\downarrow}(\mathbf{k}),\psi^-_{\uparrow}(\mathbf{k}),\psi^-_{\downarrow}(\mathbf{k})\right\}^{T} \equiv \left\{\psi_1,\psi_2,\psi_3,\psi_4\right\}^{T}
\end{equation}
and the TR conjugate part is given by
\begin{equation}
\label{9-Model-Nambu-Psi-2}
\Psi_{\Theta} = \left\{\psi^+_{\downarrow}(-\mathbf{k}),-\psi^+_{\uparrow}(-\mathbf{k}),\psi^-_{\downarrow}(-\mathbf{k}), -\psi^-_{\uparrow}(-\mathbf{k})\right\}^{\dag} \equiv \left\{\psi_5,\psi_6,\psi_7,\psi_8\right\}^{T},
\end{equation}
where the notation $\psi_i$ with $(i=\overline{1,8})$ is introduced for convenience.

It is difficult to find eigenstates of the BdG Hamiltonian (\ref{BdG-Hamiltonian}) in an explicit analytic form when $\hat{\Delta} \ne 0$. The situation changes in the case of a strong pseudomagnetic field $B_5 \to \infty$ when the approximation of the lowest pseudo-Landau level can be used. To see this, we begin with solutions $\psi^+_{\downarrow}$ for quasiparticles of chirality $\lambda=+$ [without loss of generality, we assume that $\sign{eB_5} =1$]. They are defined by the equation
\begin{equation}
\left[-i\partial_x -i\frac{eB_5}{c}x -b_x -i(-i\partial_y -b_y)\right]\psi^+_{\downarrow} = 0,
\label{chirality-plus}
\end{equation}
which gives
\begin{equation}
\psi^+_{\downarrow}=N_{+}\,e^{-\frac{1}{2l_{B_5}^2}\left[l_{B_5}^2 (k_y-b_y)+x\right]^2} e^{ik_z z+i k_y y +ib_x x},
\label{chirality-plus-solution-sB5+}
\end{equation}
where $N_{+}$ is the normalization factor and $l_{B_5}=\sqrt{c/|eB_5|}$ is the pseudomagnetic length.

For quasiparticles of negative chirality $\lambda=-$, the equation and its solution read
\begin{equation}
\left[-i\partial_x -i\frac{eB_5}{c}x +b_x +i(-i\partial_y +b_y)\right]\psi^{-}_{\uparrow} = 0
\label{chirality-minus}
\end{equation}
and
\begin{equation}
\psi^{-}_{\uparrow}=N_{-}\,e^{-\frac{1}{2l_{B_5}^2}\left[-l_{B_5}^2 (k_y+b_y)+x\right]^2} e^{ik_z z+i k_y y -ib_x x},
\label{chirality-minus-solution-sB5+}
\end{equation}
respectively. The states $\psi^+_{\downarrow}$ and $\psi^{-}_{\uparrow}$ form the basis that will be used in the analysis of the superconducting pairing below.

\subsection{Pairing and gap equation}
\label{sec:pairing-pairing}

Let us begin our analysis with the {\it intra-node} pairing of quasiparticles of chirality $\lambda=+$. Since only $\psi^+_{\downarrow}$ is not zero, the inter-node $s$-wave pairing is not possible. Indeed, the corresponding anomalous average $\langle\psi^+_{\downarrow}\psi^+_{\downarrow}\rangle$ vanishes identically due to the Pauli principle. Thus, the intra-node pairing is not permitted in the lowest pseudo-Landau level approximation.

Obviously, the situation is different for the {\it inter-node} pairing, where the nonzero anomalous average $\langle\psi^+_{\downarrow}\psi^-_{\uparrow}\rangle$ is possible. The general gap matrix for the inter-node pairing is
\begin{equation}
\hat{\Delta}_{\rm inter} = \left(
                 \begin{array}{cc}
                   0 & \left(\bm{\Delta}\cdot \bm{\sigma}\right) \\
                   -\left(\bm{\Delta}\cdot \bm{\sigma}\right) & 0 \\
                 \end{array}
               \right).
\label{internode-gap}
\end{equation}
Among the three possible types of superconducting gaps, it is $\Delta_z$ which describes the only possible $s$-wave anomalous average $\langle\psi^+_{\downarrow}\psi^-_{\uparrow}\rangle$.

\begin{figure*}[!ht]
\centering
\subfigure[]{\includegraphics[width=0.35\textwidth]{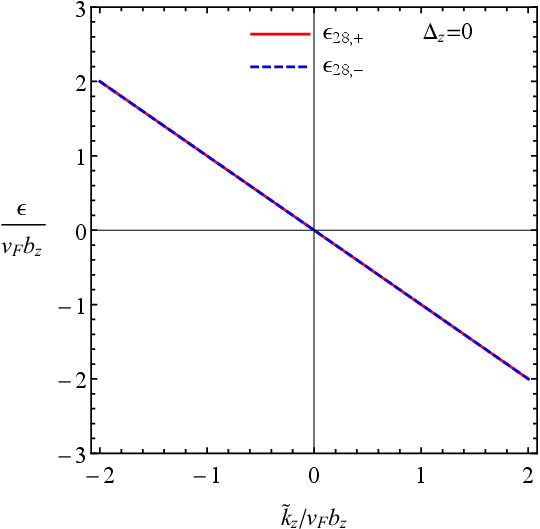}}
\hspace{0.1\textwidth}
\subfigure[]{\includegraphics[width=0.35\textwidth]{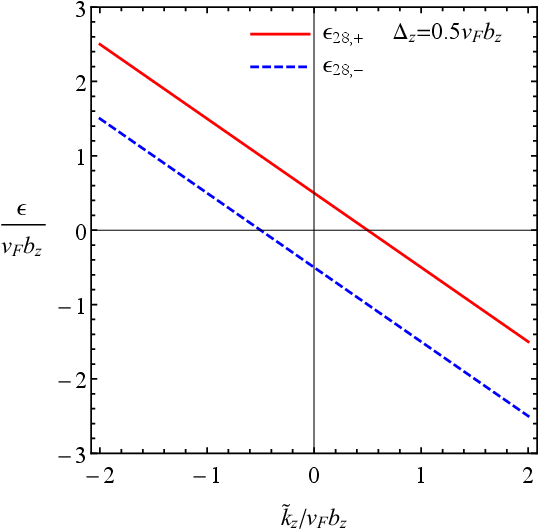}}
\caption{The energy dispersion relation $\epsilon_{28, \pm}$ (red solid and blue dashed lines) as a function of $\tilde{k}_z=v_F (k_z-b_z)$ at $\mu=0$ for $\Delta_z=0$ (panel (a)) and $\Delta_z=0.5v_F b_z$ (panel (b)).
}
\label{fig:energy-dispersion-B5}
\end{figure*}

The eigenstate equation (\ref{BdG-eigenstate}) for the inter-node pairing gives
\begin{eqnarray}
\label{internode-equation-1-be}
&&-\left[v_F (k_z-b_z)+\mu+\epsilon\right]\psi_2 - \Delta_z\psi_8=0,\\
&&-\Delta_z^{*}\psi_2+\left[-v_F\left(k_z-b_z\right)+\mu-\epsilon\right]\psi_8=0,
\label{internode-equation-1-ee}
\end{eqnarray}
and similar equations for $\psi_3$ and $\psi_5$ with $b_z\to-b_z$. Equations (\ref{internode-equation-1-be}) and (\ref{internode-equation-1-ee}) lead to the following BdG energy dispersion:
\begin{equation}
\epsilon_{28, \pm}=-v_F (k_z -b_z)\pm \sqrt{\mu^2+|\Delta_z|^2}.
\label{energy-dispersion-1}
\end{equation}
Clearly, $\epsilon_{35, \pm} = \epsilon_{28, \pm}(b_z\to-b_z)$. The energy dispersion relation $\epsilon_{28, \pm}$ is plotted in Fig.~\ref{fig:energy-dispersion-B5} for two values of $\Delta_z$. As one can see, the superconducting gap $\Delta_z$ {\it does not} open the gap in the energy spectrum. In fact, it splits the degenerate energy branches. As shown in Appendix~\ref{sec:App-B}, due to a different structure of Landau levels, the case of strong conventional magnetic field is qualitatively different. In particular, the $s$-wave inter-node pairing leads to a gapped energy spectrum (see Fig.~\ref{fig:energy-dispersion-B}).

The wave functions $\psi_2$ and $\psi_8$ are normalized as
\begin{eqnarray}
\label{wave-functions-norm-28}
\int d^3r \left\{\left[\psi_2(\mathbf{k})\right]^{*}\psi_2(\mathbf{k}^{\prime}) +\left[\psi_8(\mathbf{k})\right]^{*}\psi_8(\mathbf{k}^{\prime}) \right\}  = (2\pi)^2 \delta(k_z^{\prime}-k_z)\delta(k_y^{\prime}-k_y),
\end{eqnarray}
which gives
\begin{eqnarray}
\label{wave-functions-norm-28-1}
|N_{+}|^2 =\frac{1}{\sqrt{\pi} l_{B_5}} \left(1+ \frac{|\Delta_z|^2}{\left[\mu-\epsilon_{28,\pm} -v_F (k_z-b_z)\right]^2} \right)^{-1}.
\end{eqnarray}
Similar expressions with $b_z\to-b_z$ hold for $\psi_3$ and $\psi_5$.

By varying the effective action (\ref{effective-action}) with respect to $\Delta^{\dagger}_z$, we find the following gap equation in the lowest pseudo-Landau level approximation:
\begin{eqnarray}
\label{gap-equation-0}
-\frac{4\Delta_z}{g} - i \int \frac{d\omega\, d^2k}{(2\pi)^3} \mbox{tr}\left[\Psi_{\rm BdG}^{\dag} \tau_{-}\otimes (-i)\tau_y \otimes \sigma_z \left[\omega +i0 \sign{\omega} -H_{\rm BdG}\right]^{-1}  \Psi_{\rm BdG}
\right]=0,
\end{eqnarray}
where $\tau_{-}=(\tau_x-i\tau_y)/2$ and the first $\bm{\tau}$-matrix acts in the Nambu space. Explicitly, the gap equation reads
\begin{eqnarray}
\label{gap-equation-1}
\Delta_z = \frac{ig}{4} \int \frac{d\omega\, d^2k}{(2\pi)^3} \sum_{\pm}\left[\frac{\psi_{\downarrow}^{+}(\mathbf{k})\psi_{\uparrow}^{-}(-\mathbf{k})}{\omega+i0 \sign{\omega} -\epsilon_{28,\pm}}
-\frac{\psi_{\uparrow}^{-}(\mathbf{k}) \psi_{\downarrow}^{+}(-\mathbf{k})}{\omega+i0 \sign{\omega} -\epsilon_{35,\pm}}\right].
\end{eqnarray}
Let us consider the first term in the square brackets. By using Eqs.~(\ref{chirality-plus-solution-sB5+}), (\ref{energy-dispersion-1}), and (\ref{wave-functions-norm-28-1}), we obtain
\begin{eqnarray}
\label{gap-equation-1-first}
i\int \frac{d\omega\, d^2k}{(2\pi)^3} \sum_{\pm}\frac{\psi_{\downarrow}^{+}(\mathbf{k})\psi_{\uparrow}^{-}(-\mathbf{k})}{\omega+i0 \sign{\omega} -\epsilon_{28,\pm}} =i \int \frac{d\omega\, dk_z}{(2\pi)^3} \frac{1}{l_{B_5}^2} \frac{\Delta_z}{\left[\omega+i0 \sign{\omega}+v_F (k_z-b_z)\right]^2 -\mu^2-|\Delta_z|^2}.
\end{eqnarray}
The integrals over $\omega$ and $k_z$ can be taken straightforwardly
\begin{eqnarray}
\label{gap-equation-1-first-v1}
&&\int d\omega\, \int_{-\Lambda}^{\Lambda} dk_z  \frac{1}{\left[\omega+i0 \sign{\omega}+v_F (k_z-b_z)\right]^2 -\mu^2-|\Delta_z|^2} \nonumber\\
&&=-i \pi \int d\omega\, \int_{-\Lambda}^{\Lambda} \mbox{sign}{\left[\omega \left(\omega +v_F (k_z -b_z)\right)\right]} \delta \left[\left(\omega +v_F (k_z -b_z)\right)^2- \mu^2 -|\Delta_z|^2\right] = -\frac{2\pi i}{v_F},
\end{eqnarray}
where $\Lambda$ is a momentum cutoff, which is usually determined by the range of applicability of the low-energy linearized model. One can check that the second term in the square brackets in Eq.~(\ref{gap-equation-1}) gives the same expression albeit with an opposite sign. Therefore, it doubles the final result.

Thus, the gap equation (\ref{gap-equation-1}) reads
\begin{eqnarray}
\label{gap-equation-2-v1}
\Delta_z= \frac{g\Delta_z}{8\pi^2v_F l_{B_5}^2}.
\end{eqnarray}
It admits only a trivial solution $\Delta_z=0$. (Strictly speaking, a nontrivial solution is possible for a certain value of the coupling constant $g$ or the pseudomagnetic field $B_5$. Since it is unlikely that such a solution survives beyond the lowest pseudo-Landau level approximation, we omit it as the spurious one). Therefore, we conclude that a strong pseudomagnetic field does not allow for the intrinsic $s$-wave superconductivity in Weyl semimetals. On the other hand, there is another well-known way to realize the superconducting state through the proximity effect, which we consider in the next section. It is worth noting also that the gap generation in a strong magnetic field is qualitatively different. In particular, there is a nontrivial solution for the corresponding gap equation (see Appendix~\ref{sec:App-B} for details).

\section{Proximity effect}
\label{sec:proximity}

Superconducting gaps are routinely induced in materials, which do not support an intrinsic superconductivity, by coupling them to superconductors. In essence, this phenomenon is connected with the permeation of Cooper pairs into a nonsuperconducting medium and is known as the proximity or Holm--Meissner effect~\cite{Holm:1932}. As we argue in this paper, the proximity effect in Weyl semimetals is unusual since it is affected strongly by a pseudomagnetic field. Indeed, as we discussed before, this field is {\it always} present near the surface of a Weyl semimetal~\cite{Chernodub-Vozmediano:2014,Grushin:2016,Grushin:2018,Benito-Matias-Gonzalez:2020}. While pseudomagnetic field is, in general, nonuniform, we assume that it remains strong and changes weakly at the surface of the semimetal.

The simplest approach to the proximity effect is to add a bare superconducting gap term $\hat{\Delta}_0$ to the initial BdG Hamiltonian, which corresponds to the internode pairing channel (\ref{internode-gap}). In this case, $\hat{\Delta}_0=\tau_{+}\otimes\sigma_z\Delta_0$ and $\tau_{+}=(\tau_x+i\tau_y)/2$. Actually, the description of the proximity effect is a rather delicate issue. For example, as follows from McMillan's approach~\cite{McMillan:1968}, the induced term acquires an energy dependence. While this more complicated case will be considered in Sec.~\ref{sec:proximity-SE}, it is instructive to begin our analysis of the proximity effect in Weyl semimetals with the simplest approach, where we assume that the quasiparticle energy in a Weyl semimetal is much lower than the gap in the superconductor.

\subsection{Naive consideration}
\label{sec:proximity-B5}

\begin{figure*}[!ht]
\centering
\subfigure[]{\includegraphics[width=0.35\textwidth]{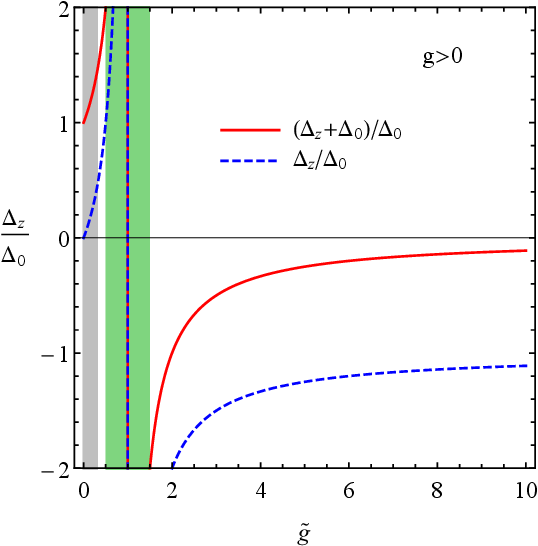}}
\hspace{0.1\textwidth}
\subfigure[]{\includegraphics[width=0.35\textwidth]{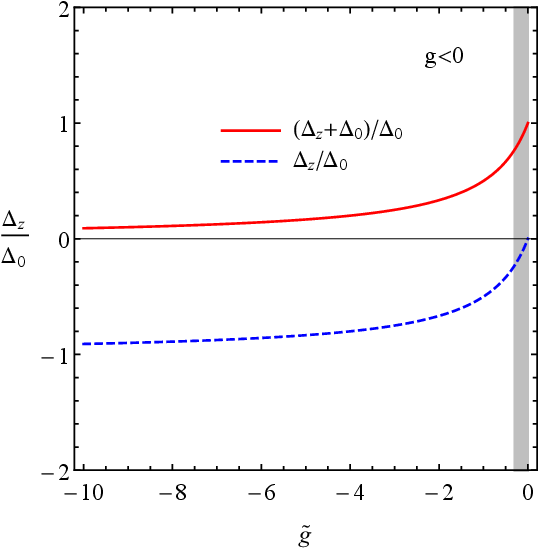}}
\caption{The dependence of the full $\Delta_z+\Delta_0$ (red solid line) and intrinsic $\Delta_z$ (blue dashed) superconducting gaps on $\tilde{g} =g |eB_5|/(8\pi^2 cv_F)$, where the coupling constant $g$ is positive for attraction and negative for repulsion. The gray areas show schematically the excluded regions of a weak pseudomagnetic field $|\mu| \leq \sqrt{2}v_F/l_{B_5}$. The green region in panel (a) corresponds to the values of the pseudomagnetic field where criterion (\ref{proximity-LLL-app}) is not satisfied.
}
\label{fig:proximity-Deltaz-B5}
\end{figure*}

By replacing $\Delta_z\to\Delta_z+\Delta_0$ on the right-hand side of Eq.~(\ref{gap-equation-2-v1}), we obtain the following gap equation:
\begin{eqnarray}
\label{proximity-gap-equation-2-v1}
\Delta_z= \tilde{g} \left(\Delta_z+\Delta_0\right),
\end{eqnarray}
where we introduced the shorthand notation for the effective interaction constant
\begin{equation}
\label{proximity-B5-tg}
\tilde{g} = \frac{g |eB_5|}{8\pi^2 cv_F},
\end{equation}
which is linear in the pseudomagnetic field strength. Unlike the study of the intrinsic superconductivity in the previous section, a nontrivial solution for $\Delta_z$ is possible now. It reads
\begin{eqnarray}
\label{proximity-Deltaz}
\Delta_z = \frac{\tilde{g}\Delta_0}{1-\tilde{g}}
\end{eqnarray}
and the full gap equals
\begin{eqnarray}
\label{proximity-Deltaz-full}
\Delta_z+\Delta_0 = \frac{\Delta_0}{1-\tilde{g}}.
\end{eqnarray}
As one can see from this expression and Fig.~\ref{fig:proximity-Deltaz-B5}, a strong pseudomagnetic field suppresses the full gap, which vanishes in the limit $|B_5|\to\infty$. It is notable that, as shown in Fig.~\ref{fig:proximity-Deltaz-B5}(a), there is an interesting regime for a positive coupling constant $g>0$ (attraction) where a pole in both induced and full gaps at a certain critical value of the pseudomagnetic field appears. According to Eq.~(\ref{proximity-Deltaz-full}) and Fig.~\ref{fig:proximity-Deltaz-B5}(b), such a pole is absent for $g<0$ (repulsion). Note, however, that the lowest pseudo-Landau level approximation where
\begin{equation}
\label{proximity-LLL-app}
\sqrt{2} \frac{v_F}{l_{B_5}} = v_F \sqrt{\frac{2|eB_5|}{c}} \gg |\Delta_z + \Delta_0|
\end{equation}
is not applicable near the pole because the gap diverges there [see the green region in Fig.~\ref{fig:proximity-Deltaz-B5}(a)]. In addition, the chemical potential in the Weyl semimetal should be sufficiently low $|\mu| \leq \sqrt{2}v_F/l_{B_5}$ that excludes the region of small $\tilde{g}$ (see the gray regions in both panels of Fig.~\ref{fig:proximity-Deltaz-B5}). As we will see in
Sec.~\ref{sec:proximity-SE}, the results for the regime of small positive $\tilde{g}<1$ qualitatively agree with the self-energy approach if one sets $\sign{\Delta_z \Delta_0}=-1$ in Eq.~(\ref{proximity-Deltaz}).

\subsection{Self-energy approach}
\label{sec:proximity-SE}

In this subsection, we provide a more rigorous description of the proximity-induced superconductivity. By using McMillan's model~\cite{McMillan:1968} (this model is routinely used in the study of the proximity effect in many physical systems, e.g., in topological insulators~\cite{Stanescu-DasSarma:2010,Potter-Lee:2011,Lutchyn-DasSarma:2012,Maier-Molenkamp:2012,Tkachov-Hankiewicz:2013,Tkachov:2013}), we calculate the proximity-induced self-energy correction to the quasiparticle propagator and the effective BdG Hamiltonian in a Weyl semimetal.

\subsubsection{Self-energy and induced terms}
\label{sec:proximity-SE-model}

The BdG Hamiltonian of a usual $s$-wave spin-singlet superconductor reads
\begin{eqnarray}
\label{PSE-2-HS-def}
\hat{H}_{\rm SC} = \left(
                \begin{array}{cc}
                  \xi_{\mathbf{k}}& \Delta  \\
                  \Delta^{*} & -\xi_{\mathbf{k}}  \\
                \end{array}
              \right)\otimes\mathds{1}_2,
\end{eqnarray}
where $\xi_{\mathbf{k}}=k^2/(2m)-\mu$, $m$ is the effective mass of electron quasiparticles, $\Delta$ is the superconducting gap, and the unit matrix denotes the spin degree of freedom.

By using the BdG Hamiltonian (\ref{PSE-2-HS-def}), it is straightforward to derive the following Green's function for quasiparticles in the superconductor:
\begin{eqnarray}
\label{PSE-2-GSR-def}
G_{S}(\omega,\mathbf{k}) = \frac{\omega \mathds{1}_2+\tau_z\xi_{\mathbf{k}} +\Delta \tau_{+} +\Delta^{*} \tau_{-}}{\omega^2-\xi_{\mathbf{k}}^2 -|\Delta|^2} \otimes\mathds{1}_2.
\end{eqnarray}

The self-energy due to tunneling between the superconductor and a Weyl semimetal reads~\cite{McMillan:1968}
\begin{eqnarray}
\label{PSE-2-Sigma}
\Sigma(\omega, \mathbf{k},\mathbf{k}_1) = \sum_{\mathbf{k}^{\prime}} \hat{T}_{\mathbf{k},\mathbf{k}^{\prime}} G_{S}(\omega,\mathbf{k}^{\prime}) \hat{T}_{\mathbf{k}^{\prime},\mathbf{k}_1}^{T},
\end{eqnarray}
where $\hat{T}_{\mathbf{k},\mathbf{k}^{\prime}}$ is a tunneling matrix.

The approximation of the lowest pseudo-Landau level significantly simplifies the structure of the effective BdG Hamiltonian and Green's function in a Weyl semimetal. Therefore, we find it useful to introduce the reduced BdG Hamiltonian in the basis $\left\{\psi_2, \psi_3, \psi_8, \psi_5\right\}$ as follows:
\begin{eqnarray}
\label{PSE-2-HW-red-def}
\hat{H}_{\rm W} = \left(
                    \begin{array}{cccc}
                       -\mu -v_F (k_z-b_z) & 0 & \Delta_z & 0 \\
                      0 & -\mu -v_F (k_z+b_z) & 0 & \Delta_z \\
                      \Delta_z^{*} & 0 & \mu -v_F (k_z-b_z) & 0 \\
                      0 & \Delta_z^{*} & 0 & \mu -v_F (k_z+b_z) \\
                    \end{array}
                  \right).
\end{eqnarray}

We consider the tunneling matrix $\hat{T}_{\mathbf{k},\mathbf{k}^{\prime}}$ in the simplest diagonal form
\begin{eqnarray}
\label{PSE-2-T-def}
\hat{T}_{\mathbf{k},\mathbf{k}^{\prime}} = t_{\mathbf{k},\mathbf{k}^{\prime}} \tau_z \otimes \mathds{1}_2   = t_{\mathbf{k},\mathbf{k}^{\prime}} \hat{T}.
\end{eqnarray}
Further, we assume that the tunneling coupling randomly fluctuates, i.e.,
\begin{equation}
\label{PSE-2-tt-correlator}
\langle t_{\mathbf{k},\mathbf{k}^{\prime}} t_{\mathbf{k}_1,\mathbf{k}_1^{\prime}}\rangle = t_0^2 \delta_{\mathbf{k}-\mathbf{k}^{\prime},-\mathbf{k}_1+\mathbf{k}_1^{\prime}}.
\end{equation}
This approximation is known as the ``rough surface" approximation, which is valid for sufficiently large contact areas~\cite{Levitov:book}. In such a case, tunneling is the most efficient at certain parts of the interface where the barrier is the smallest. [Notice that such a treatment of tunneling is similar to the treatment of disorder.] The dependence on momentum in the tunneling matrix is not important also in the case of an isotropic superconductor and at small energies~\cite{Parks:book-1,Schrieffer:book-1964}. Then the self-energy reads
\begin{eqnarray}
\label{PSE-2-Sigma-1}
\Sigma(\omega, \mathbf{k},\mathbf{k}_1) =\delta_{\mathbf{k},\mathbf{k}_1} \Sigma(\omega, \mathbf{k}) =  \delta_{\mathbf{k},\mathbf{k}_1} t_0^2 \sum_{\mathbf{k}^{\prime}} \hat{T} G^{R}_{S}(\omega,\mathbf{k}^{\prime}) \hat{T}^{T} = -\delta_{\mathbf{k},\mathbf{k}_1} i\Gamma_0 \frac{\omega \mathds{1}_2-\Delta \tau_{+} -\Delta \tau_{-}}{\sqrt{\omega^2 -|\Delta|^2}} \otimes\mathds{1}_2,
\end{eqnarray}
where $\nu_{0,S}=m\sqrt{2m\mu}/(2\pi^2)$ is the DOS of the normal state per spin and $\Gamma_0 = \pi t_0^2 \nu_{0,S}$ is the tunneling energy scale. Having obtained the self-energy, the full Green's function is determined by the Schwinger--Dyson equation
\begin{eqnarray}
\label{PSE-2-SD-eq}
G^{-1}(\omega,\mathbf{k}) = S^{-1}(\omega,\mathbf{k}) -\Sigma(\omega,\mathbf{k}).
\end{eqnarray}
This means that the proximity effect modifies the BdG Hamiltonian as follows:
\begin{eqnarray}
\label{PSE-2-Heff}
H_{\rm BdG} \to H_{\rm BdG}+\Sigma.
\end{eqnarray}

It is straightforward to see that self-energy (\ref{PSE-2-Sigma-1}) leads to the following standard renormalization of parameters in the BdG eigenequation (\ref{BdG-eigenstate}):
\begin{eqnarray}
\label{PSE-Delta}
\Delta_z \to \tilde{\Delta}_z &=& \Delta_z + \frac{i \Gamma_0 \Delta}{\sqrt{\epsilon^2-|\Delta|^2}},\\
\label{PSE-eps}
\epsilon \to \tilde{\epsilon} &=& \epsilon+ \frac{i \Gamma_0 \epsilon}{\sqrt{\epsilon^2-|\Delta|^2}}.
\end{eqnarray}
Note that if one assumes that $|\epsilon|\ll|\Delta|$, a simple replacement $\Delta_z\to\Delta_z+\Gamma_0$ immediately follows from Eq.~(\ref{PSE-Delta}). This replacement coincides with that used in Sec.~\ref{sec:proximity-B5} if the bare gap $\Delta_0$ in the naive approach is identified with the tunneling energy scale, $\Delta_0=\Gamma_0$. Therefore, the naive approach works at low energies.

\subsubsection{Energy spectrum}
\label{sec:PSE-spectrum}

The modification of parameters in the BdG Hamiltonian defined in Eqs.~(\ref{PSE-Delta}) and (\ref{PSE-eps}) significantly changes, in general, the energy spectrum of a Weyl semimetal. Obviously, Eqs.~(\ref{internode-equation-1-be}) and (\ref{internode-equation-1-ee}) retain their form
\begin{eqnarray}
&&-\left[v_F (k_z-b_z)+\mu+\tilde{\epsilon}\right]\psi_2 - \tilde{\Delta}_z\psi_8=0,\\
&&-\tilde{\Delta}_z^{*}\psi_2+\left[-v_F\left(k_z-b_z\right)+\mu-\tilde{\epsilon}\right]\psi_8=0
\label{PSE-internode-equation-1}
\end{eqnarray}
with replacements (\ref{PSE-Delta}) and (\ref{PSE-eps}). In addition, there is a similar system for $\psi_3$ and $\psi_5$ with $b_z\to-b_z$. The energy spectrum is determined by setting the corresponding determinant to zero, i.e.,
\begin{eqnarray}
\left[\tilde{\epsilon}+v_F (k_z-b_z)\right]^2 -\mu^2 -|\tilde{\Delta}_z|^2=0.
\label{PSE-internode-equation-1-det}
\end{eqnarray}
Unfortunately, solutions to the characteristic equation (\ref{PSE-internode-equation-1-det}) can be obtained only numerically.

\begin{figure*}[!ht]
\centering
\subfigure[]{\includegraphics[width=0.35\textwidth]{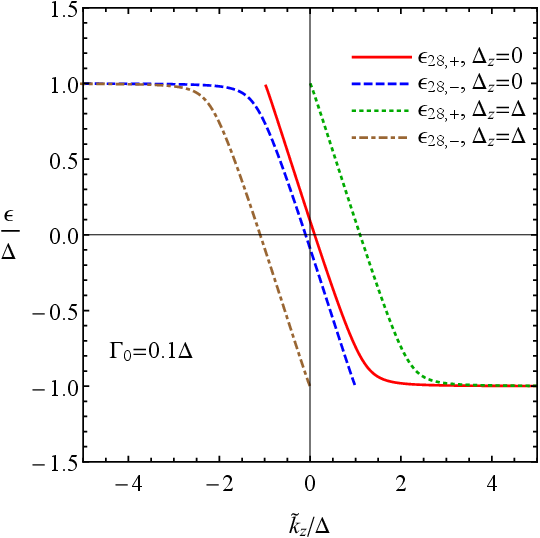}}
\hspace{0.1\textwidth}
\subfigure[]{\includegraphics[width=0.35\textwidth]{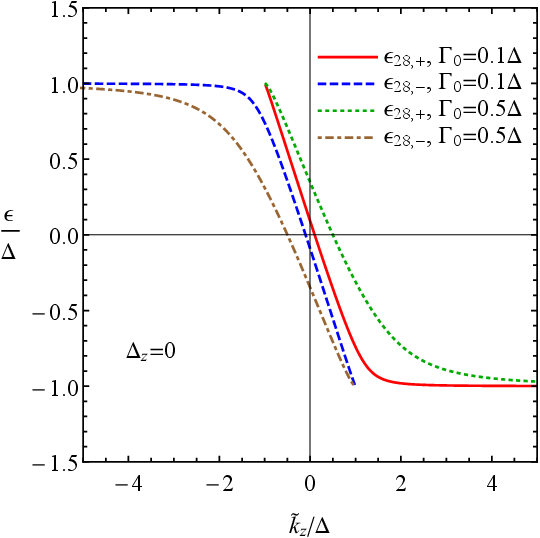}}
\caption{The energy dispersion $\epsilon_{28, \pm}$ as a function of $\tilde{k}=v_F (k_z-b_z)$ at fixed $\Gamma_0=0.1\Delta$ and a few values of $\Delta_z$ (panel (a)) as well as $\Delta_z=0$ and a few values of $\Gamma_0$ (panel (b)). In both panels, $\mu=0$.
}
\label{fig:PSE-eps-28-filtered-B5}
\end{figure*}

We present the numerical results for the energy spectrum in a Weyl semimetal in Figs.~\ref{fig:PSE-eps-28-filtered-B5}(a) and \ref{fig:PSE-eps-28-filtered-B5}(b) for a few values of $\Delta_z$ and $\Gamma_0$, respectively. As one can see from Fig.~\ref{fig:PSE-eps-28-filtered-B5}(a), $\Delta_z$ enhances the separation of the two energy branches but it does not alter other features of the spectrum. On the other hand, according to Fig.~\ref{fig:PSE-eps-28-filtered-B5}(b), the role of $\Gamma_0$ is to enhance the separation and flatten the step-like features. Furthermore, like the surface plasmon-polaritons in Weyl semimetals~\cite{Hofmann-DasSarma:2016,Kotov-Lozovik:2018,Tamaya-Kawabata:2019}, the
energy dispersions are nonreciprocal and end abruptly. Such an abrupt change is similar to the interface bound states in superconductor-graphene junctions~\cite{Burset-Levy-Yeyati:2009,Casas-Herrera:2019}, where, however, nonreciprocity was absent. Therefore, we can argue that the nonreciprocal spectrum shown in Fig.~\ref{fig:PSE-eps-28-filtered-B5} is a characteristic feature of the proximity effect in Weyl semimetals in strong pseudomagnetic fields.

To provide an analytical insight into the nonreciprocity, let us consider the characteristic equation (\ref{PSE-internode-equation-1-det}) in the limit $\epsilon \to -\Delta$ and $\tilde{k}_z\to \pm\infty$. It takes the following form:
\begin{equation}
\label{PSE-char-eq-limit}
\left(\epsilon^2 -|\Delta|^2\right) \alpha^2 +2\alpha \epsilon\left(\epsilon +\tilde{k}_z\right) +\left(\epsilon+\tilde{k}_z\right)^2 -\mu^2=0,
\end{equation}
where $\alpha=\Gamma_0/\sqrt{|\Delta|^2-\epsilon^2}$. The first term in the above equation gives constant $-\Gamma_0^2$. The second term diverges at $\epsilon \to -\Delta$ and $\tilde{k}_z\to \pm\infty$. To satisfy the equation, this term should be compensated by the third term which diverges also at $\tilde{k}_z\to \pm\infty$. However, since these terms have the same sign, no cancellation is possible at $\tilde{k}_z\to -\infty$ leading to a nonreciprocal dependence on momentum.

\subsubsection{Gap equation}
\label{sec:PSE-gap}

Now let us study how the proximity-induced corrections affect the dynamical generation of $\Delta_z$ in a Weyl semimetal. The gap equation (\ref{gap-equation-1}) has the same form where one should use the energy spectrum defined by Eq.~(\ref{PSE-internode-equation-1-det}). By performing straightforward algebraic manipulations and taking integral over $\omega$, we obtain
\begin{eqnarray}
\label{PSE-gap-equation-3}
\Delta_z = \tilde{g} I(\Delta_z),
\end{eqnarray}
where
\begin{eqnarray}
\label{PSE-gap-equation-gint-I}
\quad I(\Delta_z) = \int d\tilde{k}_z \frac{\tilde{\Delta}_z(\epsilon_{28,+})}{\sqrt{\mu^2 +|\tilde{\Delta}_z(\epsilon_{28,+})|^2}} \sign{\epsilon_{28,+}}.
\end{eqnarray}
The integral over $\tilde{k}_z$ is divergent for the energy spectrum shown in Fig.~\ref{fig:PSE-eps-28-filtered-B5} and a cutoff $\Lambda$ should be introduced. For our numerical calculations, we use $\Lambda =10^2\Delta$.

Let us provide an analytical estimation of the integral in Eq.~(\ref{PSE-gap-equation-gint-I}). We assume that the energy spectrum $\epsilon_{28,+}$ at small $\Gamma_0$ can be approximated as follows
\begin{equation}
\label{PSE-gap-equation-eps-28-app}
\epsilon_{28,+} \approx -\tilde{k}_z\theta\left(\tilde{k}_z +|\tilde{k}_{0}|\right)\theta\left(2\Delta-|\tilde{k}_{0}|-\tilde{k}_z\right) -(\Delta-\delta) \theta\left(\tilde{k}_z -2\Delta+|\tilde{k}_0|\right),
\end{equation}
where $\tilde{k}_{0} \simeq \Delta_z-\Delta$ denotes the termination point in the spectrum and $\delta\to+0$. Then, by setting $\mu=0$, $I(\Delta_z)$ reads
\begin{eqnarray}
\label{PSE-gap-equation-gint-I-1}
\quad I(\Delta_z) \approx -\int_{-|\tilde{k}_0|}^{2\Delta-|\tilde{k}_0|} d\tilde{k}_z \sign{\tilde{k}_z}\sign{\Delta_z +\frac{\Gamma_0 \Delta}{\sqrt{|\Delta|^2-\tilde{k}_z^2}}}
-\int_{2\Delta-|\tilde{k}_0|}^{\Lambda} d\tilde{k}_z \sign{\Delta}\sign{\Delta_z +\frac{\Gamma_0 \Delta}{\delta}}.
\end{eqnarray}
Since the last term is linearly divergent, it is determined by the cutoff and can attain a large value. Therefore, we can neglect the first term. Then
\begin{eqnarray}
\label{PSE-gap-equation-gint-I-2}
\quad I(\Delta_z) \approx  -\sign{\Delta}\sign{\Delta_z +\frac{\Gamma_0 \Delta}{\delta}} \left(\Lambda -2\Delta+|\tilde{k}_{0}|\right) \approx -\left(\Lambda -|\tilde{k}_{0}|\right) \approx -(\Lambda-\Delta) +\Delta_z.
\end{eqnarray}

Numerical results are in good agreement with the simple estimate given in Eq.~(\ref{PSE-gap-equation-gint-I-2}). In particular, we found that, for $\Lambda =10^2\Delta$ and $\Gamma_0=0.1\Delta$, the function $I(\Delta_z)$ reads
\begin{eqnarray}
\label{PSE-gap-equation-I-Deltaz}
I(\Delta_z) \approx -98.8\Delta +\Delta_z.
\end{eqnarray}
It is clear that the dependence of the right-hand side of the gap equation on $\Delta_z$ is weak. Nevertheless, it agrees qualitatively with the results obtained in the simple approximation in Sec.~\ref{sec:proximity-B5} at least when $\Delta_z$ is much smaller than the cutoff. It is important to emphasize that since $I(0)\neq0$, there is no trivial solution $\Delta_z=0$. The dependence of $\Delta_z$ on $\tilde{g}$ is well fitted by the following expression [cf. with Eq.~(\ref{proximity-Deltaz})]:
\begin{eqnarray}
\label{PSE-gap-equation-Deltaz-tg}
\Delta_z \approx -98.8\frac{\tilde{g}\Delta}{1-\tilde{g}} \approx -97.8\tilde{g} \Delta.
\end{eqnarray}
As one can see, $|\Delta_z|$ grows with the absolute value of the effective interaction constant $\tilde{g}$. This resembles the growth of $|\Delta_z|$ shown in Fig.~\ref{fig:proximity-Deltaz-B5} at small $\tilde{g}$.

\section{Density of states, spectral function, and tunneling current}
\label{sec:observables}

Let us discuss now how the unique character of proximity-induced superconductivity in a Weyl semimetal caused by the presence of strong pseudomagnetic fields could be observed experimentally. For this, we consider such quantities as the electron DOS, the spectral function, and the tunneling current.

\subsection{Density of states}
\label{sec:obs-DOS}

Let us begin with the electron DOS for Weyl semimetal, which is defined as
\begin{eqnarray}
\label{obs-DOS-nuWR-def}
\nu_{W} &=& -\frac{1}{\pi} \mbox{Im}\int \frac{dk_zdk_y}{(2\pi)^2} \mbox{tr} \left[\frac{\mathds{1}_2+\tau_z}{2}G_{W}^{R}\left(\omega, \mathbf{k}\right) \right]  \nonumber\\
&=& -\frac{1}{\pi} \mbox{Im}\int \frac{dk_zdk_y}{(2\pi)^2} \sum_{\pm}\left\{
\frac{|N_{+}|^2e^{-\frac{1}{l_{B_5}^2}\left[l_{B_5}^2 (k_y -b_y) +x\right]^2}}{\omega +i 0 -\epsilon_{28,\pm}} +\frac{|N_{-}|^2e^{-\frac{1}{l_{B_5}^2}\left[l_{B_5}^2 (k_y +b_y) +x\right]^2}}{\omega +i 0 -\epsilon_{35,\pm}}\right\},
\end{eqnarray}
where $|N_{+}|$ is defined in Eq.~(\ref{wave-functions-norm-28-1}). It suffices to consider only the first term because the second gives a similar contribution with the replacement $b_z\to -b_z$. By integrating over $k_y$, we obtain
\begin{eqnarray}
\label{obs-DOS-nuWR-first}
&&-\frac{1}{\pi  l_{B_5}^2} \mbox{Im}\int \frac{dk_z}{(2\pi)^2} \sum_{\pm} \frac{1}{2} \frac{1}{\omega +i 0 -\epsilon_{28,\pm}} \nonumber\\
&&= -\frac{1}{\pi v_F l_{B_5}^2} \mbox{Im}\int \frac{d\tilde{k}_z}{(2\pi)^2} \frac{1}{2}\left[\frac{1}{\omega +i0 -\epsilon_{28,+}(\tilde{k}_z)} +\frac{1}{\omega +i0 -\epsilon_{28,-}(-\tilde{k}_z)}  \right]
\nonumber\\
&&= -\frac{1}{\pi v_F l_{B_5}^2} \mbox{Im}\int \frac{d\tilde{k}_z}{(2\pi)^2} \frac{\omega}{\omega^2 +i0\sign{\omega} -\epsilon_{28,+}^2},
\end{eqnarray}
where we replaced $\tilde{k}_z\to-\tilde{k}_z$ in the second line and used $\epsilon_{28,+}(\tilde{k}_z)=-\epsilon_{28,-}(-\tilde{k}_z)$. Thus, by taking into account the contributions from both Weyl nodes, the DOS reads
\begin{equation}
\label{obs-DOS-nuWR-fin}
\nu_{W} = \frac{2}{v_F l_{B_5}^2} \int \frac{d\tilde{k}_z}{(2\pi)^2} |\omega| \delta\left(\omega^2-\epsilon_{28,+}^2\right).
\end{equation}
In general, one should use an energy spectrum defined in Eq.~(\ref{PSE-internode-equation-1-det}) and perform the integration over $\tilde{k}_z$ numerically. If the proximity effect is absent, i.e., $\Gamma_0=0$, then the integral over $\tilde{k}_z$ can be trivially taken resulting in
\begin{equation}
\label{obs-nuWR-Gamma0=0}
\nu_{0,W} = \frac{1}{2\pi^2 v_F l_{B_5}^2}.
\end{equation}
Similar to the scaling of the DOS in the lowest Landau level, the DOS in the lowest pseudo-Landau level (\ref{obs-nuWR-Gamma0=0}) scales linearly with $B_5$.

For comparison, the DOS for a usual superconductor described by Hamiltonian (\ref{PSE-2-HS-def}) is
\begin{eqnarray}
\label{obs-nuSR}
\nu_{SC} &=& -\frac{1}{\pi} \mbox{Im}\int\frac{d^3k}{(2\pi)^3} \mbox{tr}\left[\frac{\mathds{1}_2+\tau_z}{2} G^{\rm R}_{\rm S}(\omega,\mathbf{k})\right] = \frac{1}{\pi} \int \frac{d^3k}{(2\pi)^3} 2\pi \delta\left(\omega^2-\xi_{\mathbf{k}}^2 -|\Delta|^2\right) \sign{\omega} (\omega+\xi_{\mathbf{k}}) \nonumber\\
&\approx& 2\nu_{0,S} \int_{-\infty}^{\infty} d\xi \frac{\sign{\omega} (\omega +\xi)}{2\sqrt{\omega^2-|\Delta|^2}} \left[\delta\left(\xi +\sqrt{\omega^2 -|\Delta|^2}\right) +\delta\left(\xi -\sqrt{\omega^2 -|\Delta|^2}\right) \right] 
=2\nu_{0,S} \frac{|\omega|}{\sqrt{\omega^2-|\Delta|^2}} \theta\left(|\omega|-|\Delta|\right). \nonumber\\
\end{eqnarray}

\begin{figure*}[!ht]
\centering
\includegraphics[width=0.35\textwidth]{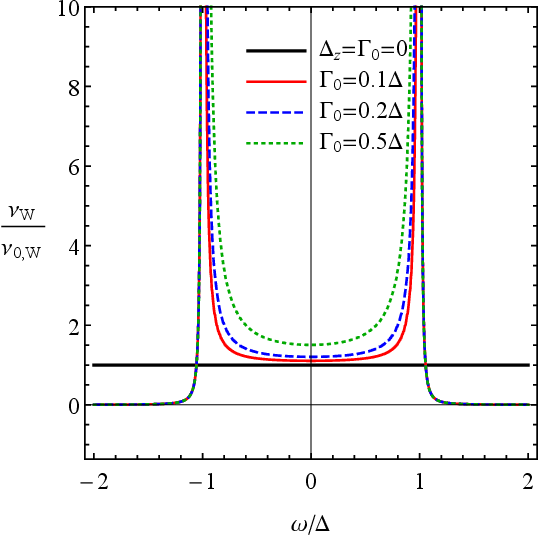}
\caption{The normalized DOS in a Weyl semimetal given in Eq.~(\ref{obs-DOS-nuWR-fin}) as a function of the frequency $\omega$ for a few values of tunneling energy scale $\Gamma_0$.}
\label{fig:obs-DOS}
\end{figure*}

We show the dependence of the DOS in a Weyl semimetal defined by Eq.~(\ref{obs-DOS-nuWR-fin}) in Fig.~\ref{fig:obs-DOS} for a few values of the tunneling energy scale $\Gamma_0$. As one can see, the proximity effect leads to noticeable peaks at $|\omega|=|\Delta|$. Moreover, while the DOS quickly vanishes at $|\omega|>|\Delta|$, it is enhanced with respect to $\nu_{0,W}$ at $|\omega|<|\Delta|$. The magnitude of enhancement is determined by the tunneling energy scale $\Gamma_0$. The dependence on $\Delta_z$ is negligible. Finally, we note that the overall scale of the DOS is dictated by the pseudomagnetic field strength
$\nu_W\sim |B_5|$.

\subsection{Spectral function}
\label{sec:obs-A}

Next, we present the results for the spectral function, which is relevant for spectroscopic studies. The spectral function could be, in principle, probed via the high-energy ARPES if the conventional superconductor is of a sufficiently small thickness. We use the standard definition of the spectral function (see also Appendix~\ref{sec:App-1})
\begin{eqnarray}
\label{tc-A-def}
A(\omega, \mathbf{k}) = - \frac{1}{\pi} \mbox{Im}\left[G_W^{\rm R}(\omega, \mathbf{k})\right]_{\mu=0}.
\end{eqnarray}
Its trace integrated over $k_y$ reads
\begin{eqnarray}
\label{tc-A-tr}
\tilde{A}(\omega,k_z)=\int dk_y\mbox{tr}\left[A(\omega, \mathbf{k})\right] = - \frac{1}{2\pi l_{B_5}^2} \sum_{\pm}\mbox{Im}\left[\frac{1}{\omega+i\delta -\epsilon_{28,\pm}} +\frac{1}{\omega+i\delta -\epsilon_{35,\pm}}\right],
\end{eqnarray}
where $\delta\to +0$. In our numerical calculations, however, we keep $\delta$ finite, which leads to a finite width of the spectral lines.

\begin{figure*}[!ht]
\centering
\subfigure[]{\includegraphics[width=0.35\textwidth]{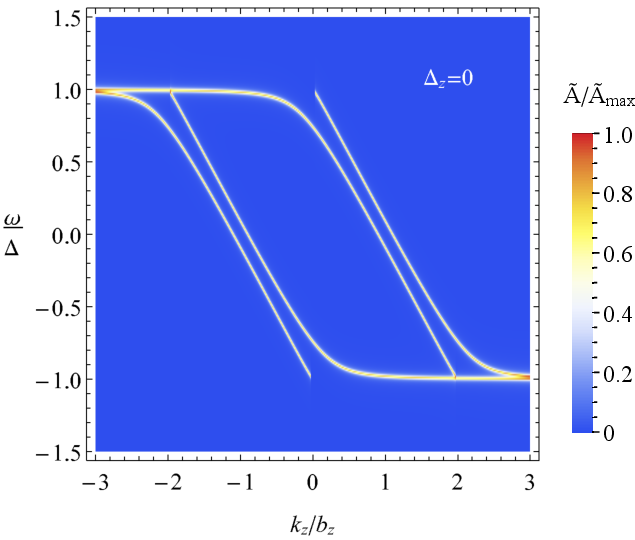}}
\hspace{0.1\textwidth}
\subfigure[]{\includegraphics[width=0.35\textwidth]{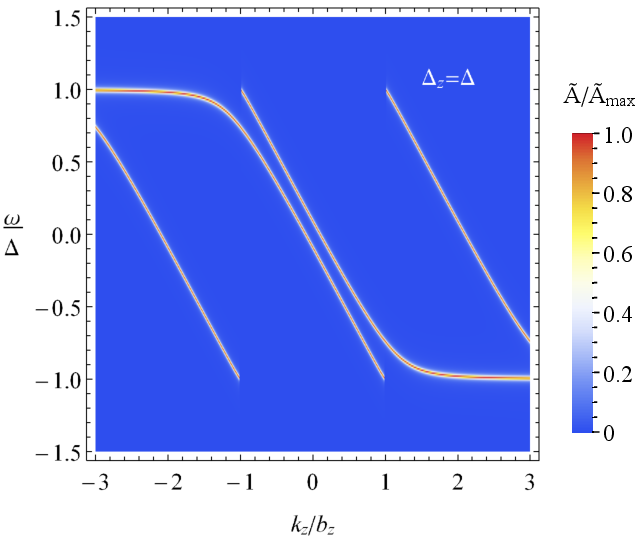}}
\caption{The dependence of the trace of the spectral function (\ref{tc-A-tr}) on the momentum component $k_z$ and the frequency $\omega$ at $\Delta_z=0$ (panel (a)) and $\Delta_z=\Delta$ (panel (b)). In both panels $\Gamma_0=0.1\Delta$, $\delta=0.01\Delta$, and $v_Fb_z=\Delta$.
}
\label{fig:obs-A}
\end{figure*}

We present the trace of the spectral function integrated over $k_y$ in Fig.~\ref{fig:obs-A}. As expected from the analysis in Sec.~\ref{sec:PSE-spectrum}, the spectral function reveals nonreciprocal branches with abrupt ends. Furthermore, since the momentum $k_z$ (rather than $\tilde{k}_z$) is used, the spectral lines corresponding to the separated Weyl nodes overlap and form a hysteresis-like curve at small $k_z$ and $\Delta_z$.

\subsection{Tunneling current}
\label{sec:tunneling-current}

A direct way to probe the proximity effect in various heterostructures is the electron tunneling. The corresponding tunneling current is sensitive to the details of the DOS in a superconducting metal and a Weyl semimetal. The current is defined by~\cite{Levitov:book}
\begin{eqnarray}
\label{tc-IV-def}
I(V) = \pi e \sum_{\mathbf{k},\mathbf{k}^{\prime}} \int d\omega\, \mbox{tr}\left[ \hat{T}_{\mathbf{k},\mathbf{k}^{\prime}}^{T} A_{W}\left(\omega +eV, \mathbf{k}^{\prime}\right) T_{\mathbf{k}^{\prime},\mathbf{k}} A_{S}\left(\omega,\mathbf{k}\right) \right] \left[n_{F}(\omega) -n_{F}(\omega+eV)\right],
\end{eqnarray}
where $V$ is an electric potential applied to the junction and the spectral function is given in Eq.~(\ref{tc-A-def}).

By using the explicit matrix structure of Green's function (\ref{PSE-2-GSR-def}), the standard definition in a Weyl semimetal,
\begin{equation}
\label{tc-G-Weyl-def}
G_{W}^{R} = \sum_{j} \frac{\psi_j \psi^{\dag}_j}{\omega +i0 -\epsilon_j},
\end{equation}
as well as Eqs.~(\ref{PSE-2-T-def}) and (\ref{PSE-2-tt-correlator}), we rewrite Eq.~(\ref{tc-IV-def}) as
\begin{eqnarray}
\label{tc-IV-expl-MB}
I(V) &=& \pi e t_0^2\sum_{\mathbf{k},\mathbf{k}^{\prime}} \int d\omega\, \mbox{tr}\Bigg\{
A_{S}\left(\omega +eV, \mathbf{k}\right)  \left[A_{W,28}\left(\omega, \mathbf{k}^{\prime}\right) +A_{W,35}\left(\omega, \mathbf{k}^{\prime}\right)\right]
+\bar{A}_{S}\left(\omega +eV, \mathbf{k}\right)  \left[\bar{A}_{W,28}\left(\omega, \mathbf{k}^{\prime}\right) +\bar{A}_{W,35}\left(\omega, \mathbf{k}^{\prime}\right)\right] \nonumber\\
&+&C_{S}\left(\omega +eV, \mathbf{k}\right)  \left[C_{W,28}\left(\omega, \mathbf{k}^{\prime}\right) +C_{W,35}\left(\omega, \mathbf{k}^{\prime}\right)\right]
+C_{S}^{\dag}\left(\omega +eV, \mathbf{k}\right)  \left[C_{W,28}^{\dag}\left(\omega, \mathbf{k}^{\prime}\right) +C_{W,35}^{\dag}\left(\omega, \mathbf{k}^{\prime}\right)\right] \Bigg\} \nonumber\\
&\times&\left[n_{F}(\omega) -n_{F}(\omega+eV)\right].
\end{eqnarray}
Here $A_{S/W}$, $\bar{A}_{S/W}$, and $C_{S/W}$ are the spectral function components defined in Appendix~\ref{sec:App-1}. As one can see, there are one-particle and Josephson terms given by the first two and last two terms in the curly brackets in Eq.~(\ref{tc-IV-expl-MB}).

In what follows, we focus on the one-particle contribution to the current. By using Eqs.~(\ref{obs-DOS-nuWR-fin}), (\ref{obs-nuSR}), (\ref{tc-A-def}), (\ref{tc-IV-expl-MB}), and expressions in Appendix~\ref{sec:App-1}, we obtain
\begin{eqnarray}
\label{tc-IV-calc}
I(V)= 2\pi e |t_0|^2 \int_{-\infty}^{\infty}d\omega\, \nu_{W} \nu_{0,S} \frac{|\omega|}{\sqrt{\omega^2-|\Delta|^2}} \theta\left(|\omega|-|\Delta|\right) \left[n_{F}(\omega) -n_{F}(\omega+eV)\right].
\end{eqnarray}
In the limit of $T\to0$, it is straightforward to derive
\begin{eqnarray}
\label{tc-IV-T0}
I(V)= 2\pi \sign{eV}  e |t_0|^2 \nu_{W} \nu_{0,S} \sqrt{e^2V^2 -|\Delta|^2} \theta(|eV|-|\Delta|).
\end{eqnarray}
As expected, the tunneling current is proportional to the density of states in both superconductor and Weyl semimetal. The DOS in a Weyl semimetal scales linearly with the strength of the pseudomagnetic field $B_5$, therefore, providing a definite signature of a strain-induced pseudomagnetic field in proximity setups.

\section{Summary}
\label{sec:Summary}

Superconducting pairing in Weyl semimetals with broken time-reversal symmetry in a strong pseudomagnetic field is investigated. Although naive arguments based on analogy with superconducting pairing in a strong magnetic field suggest that a strain-induced pseudomagnetic field, for which the Meissner effect is absent, should favor the superconducting state in Weyl semimetals, both the study in a weak field~\cite{Gorbar:2018pit} and the investigation in the ultraquantum regime performed in this paper show that the pseudomagnetic field suppresses intrinsic superconductivity. In particular, we found that only the inter-node $s$-wave pairing with the superconducting gap $\Delta_z$ is possible in the lowest pseudo-Landau level approximation. The corresponding parameter, however, does not open a gap in the energy spectrum. Its role is to split the degenerate energy branches into the two linearly dispersing ones as is evident from Fig.~\ref{fig:energy-dispersion-B5}. This behavior is drastically different from the role of an $s$-wave superconducting gap in the case of conventional Landau levels in a strong magnetic field. Indeed, the corresponding superconducting pairing does open a true gap in the energy spectrum (see Fig.~\ref{fig:energy-dispersion-B} in Appendix~\ref{sec:App-B}). By using a model with a local four-fermion interaction, we derived the gap equation and showed that it admits only a trivial solution. Therefore, unlike magnetic field, the pseudomagnetic field one does not catalyze the formation of a superconducting state in Weyl semimetals.

On the other hand, the pseudomagnetic field in a Weyl semimetal affects nontrivially the proximity effect with a conventional $s$-wave spin-singlet superconductor. In a simplified low-energy approach, the proximity effect induces the bare superconducting gap $\Delta_0$ in a Weyl semimetal. Therefore, the gap equation no longer admits a trivial solution. We found that the full superconducting gap $\Delta_z + \Delta_0$ inversely depends on the pseudomagnetic field strength $B_5$ and vanishes in the limit $B_5 \to \infty$. It is interesting that the dependence of the gap on the field strength is nonmonotonic for an attractive interaction. In particular, the magnitude of the superconducting gap grows at small fields, changes sign, and then vanishes at large fields. In the case of a repulsive interaction, $\Delta_z + \Delta_0$ decreases monotonically.

In a more refined approach where the proximity effect is taken into account via the self-energy contribution, a few important differences compared to the simple approach are found. They are well manifested at energies that are larger than a gap in a normal superconductor, where a nonreciprocal dependence on momentum and plateau-like behavior are observed. In the vicinity of the Weyl nodes, on the other hand, the energy spectrum is qualitatively the same as in the simplified approach. It is interesting that the solution to the gap equation also resembles its counterpart in the simplified approach when the coupling constant is small.

As observable signatures of the interplay of the pseudomagnetic field and superconductivity in Weyl semimetals, we propose the electron DOS, the spectral function, and the tunneling current. It is found that the electron DOS is insensitive to the superconducting gap $\Delta_z$ but has large peaks for energies close to the gap in the $s$-wave superconductor. Moreover, the DOS scales linearly with the pseudomagnetic field strength. Therefore, we believe that the tunneling current through the superconductor-Weyl semimetal interface could be an efficient means to determine the magnitude of a strain-induced pseudomagnetic field, which could be particularly large at the interface. In addition, it can be controlled by applying an external strain to a Weyl semimetal. Finally, the spectral function shows a characteristic hysteresis-like pattern composed of overlapping nonreciprocal branches. It could be probed via the ARPES when the conventional superconductor is sufficiently thin.

While in the present study we consider only the one-particle tunneling current, it would be very interesting to investigate the manifestation of the pseudomagnetic field in the Josephson current. The corresponding study will be reported elsewhere. It is worth noting also that while a pseudomagnetic field is, in general, nonuniform near the surface of a semimetal, we treated it as constant. Phenomenologically, a nontrivial spatial dependence of the pseudomagnetic field could be described by replacing $B_5\to B_5(x)$. This approximation is valid as long as the proximity-induced gap decreases in the bulk of a Weyl semimetal much faster than the field changes. A more rigorous analysis of the proximity effects is, however, beyond the scope of this study.

\begin{acknowledgments}
We are grateful to I.~A.~Shovkovy for useful comments. P.O.S. was supported by the VILLUM FONDEN via the Centre of Excellence for Dirac Materials (Grant No.~11744), the European Research Council under the European Unions Seventh Framework Program Synergy HERO, and the Knut and
Alice Wallenberg Foundation KAW 2018.0104. The work of E.V.G. was supported partially by the National Academy of Sciences of Ukraine grants No.~0116U003191 and No.~0120U100858.
\end{acknowledgments}

\appendix

\section{Spectrum and gap generation in strong magnetic field}
\label{sec:App-B}

It is instructive to consider the superconducting pairing in a strong magnetic field. Similarly to Sec.~\ref{sec:pairing-model}, we consider the minimal model of Weyl semimetal with two Weyl nodes of opposite chiralities separated by $2\mathbf{b}$ in momentum space. The linearized Hamiltonian is given by Eq.~(\ref{9-Model-hatH}) where
\begin{equation}
\label{App-B-H-chi}
H_{\lambda}=-\mu +\lambda v_F \bm{\sigma} \cdot\left(-i\bm{\nabla} + \frac{e}{c} \mathbf{A} -\lambda \mathbf{b}\right).
\end{equation}
Here $\lambda=\pm$ is the chirality of Weyl nodes, $\mu$ is the electric chemical potential, $v_F$ is the Fermi velocity, $\bm{\sigma}$ is the vector of the Pauli matrices, $c$ is the speed of light, and $\mathbf{A}=Bx\hat{\mathbf{y}}$ is the gauge field. Without the loss of generality, we assume that $\sign{eB} =1$. Nontrivial solutions for the wave functions are
\begin{equation}
\psi^+_{\downarrow}=N_{+}\,e^{-\frac{1}{2l_{B}^2}\left[l_{B}^2 (k_y-b_y)+x\right]^2} e^{ik_z z+i k_y y +ib_x x}
\label{App-B-chirality-plus-solution}
\end{equation}
and
\begin{equation}
\psi^{-}_{\downarrow}=N_{-}\,e^{-\frac{1}{2l_{B}^2}\left[l_{B}^2 (k_y+b_y)+x\right]^2} e^{ik_z z+i k_y y -ib_x x},
\label{App-B-chirality-minus-solution}
\end{equation}
where $N_{\pm}$ is the normalization factor and $l_{B}=\sqrt{c/|eB|}$ is the magnetic length. There are two nonvanishing components of  the Bogolyubov–de Gennes (BdG) wave functions [see Eqs.~(\ref{9-Model-Nambu-Psi-1}) and (\ref{9-Model-Nambu-Psi-2})], i.e., $\psi^+_{\downarrow}$ and $\psi^-_{\downarrow}$. Therefore, only $\Delta_x$ and $\Delta_y$ gaps should be considered, i.e., $\bm{\Delta}=(\Delta_x, \Delta_y,0)$.

The eigenstate equation (\ref{BdG-eigenstate}) for the inter-node pairing gives
\begin{eqnarray}
&&-\left[v_F (k_z-b_z)+\mu+\epsilon\right]\psi_2 +\left(\Delta_x+i\Delta_y\right)\psi_7=0,\\
&&\left(\Delta_x+i\Delta_y\right)^{*}\psi_2+\left[v_F\left(k_z-b_z\right)+\mu-\epsilon\right]\psi_7=0
\label{App-B-internode-equation-1}
\end{eqnarray}
and similar equations for $\psi_4$ and $\psi_5$ with $b_z\to-b_z$. The energy dispersion relations are
\begin{eqnarray}
\label{App-B-energy-dispersion-1}
\epsilon_{27, \pm}&=&\pm \sqrt{\left[\mu+ v_F (k_z -b_z)\right]^2 +|\Delta_x|^2+|\Delta_y|^2},\\
\label{App-B-energy-dispersion-2}
\epsilon_{45, \pm}&=& \epsilon_{27, \pm}(b_z\to -b_z).
\end{eqnarray}

\begin{figure*}[!ht]
\centering
\subfigure[]{\includegraphics[width=0.35\textwidth]{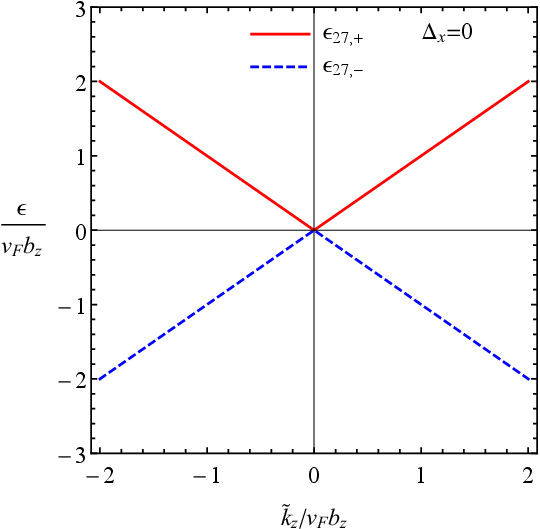}}
\hspace{0.1\textwidth}
\subfigure[]{\includegraphics[width=0.35\textwidth]{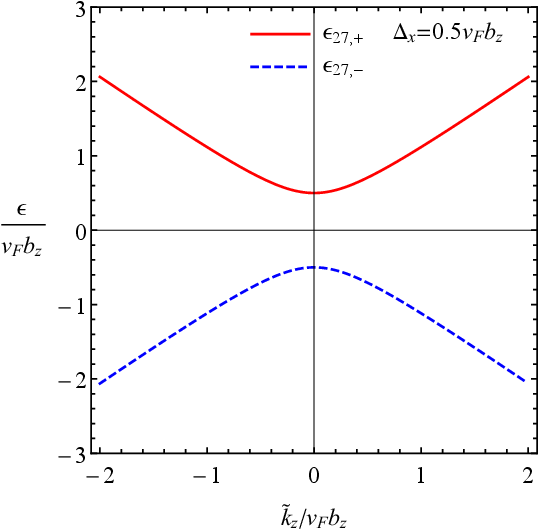}}
\caption{The dependence of the energy dispersion relation $\epsilon_{27, \pm}$ (red solid and blue dashed lines) on $\tilde{k}_z=v_F (k_z-b_z)$ at $\mu=0$ for $\Delta_x=0$ (panel (a)) and $\Delta_x=0.5v_F b_z$ (panel (b)).
}
\label{fig:energy-dispersion-B}
\end{figure*}

We present the energy dispersion relation (\ref{App-B-energy-dispersion-1}) in Fig.~\ref{fig:energy-dispersion-B}. Unlike the spectrum in the pseudomagnetic field $B_5$ plotted in Fig.~\ref{fig:energy-dispersion-B5}, it is clear that $\Delta_x$ and $\Delta_y$ open a gap in the energy spectrum.

Further, we discuss the solutions to the gap equation. For the sake of definiteness, we set $\Delta_y=0$. By varying the effective action (\ref{effective-action}) with respect to $\Delta^{\dagger}_x$, we obtain
\begin{eqnarray}
\label{App-B-gap-equation-1}
\Delta_x &=& \frac{ig}{4} \int \frac{d\omega\, d^2k}{(2\pi)^3} \sum_{\pm}\left[\frac{\psi_{\downarrow}^{+}(\mathbf{k})\psi_{\downarrow}^{-}(-\mathbf{k})}{\omega+i0 \sign{\omega} -\epsilon_{27,\pm}}
-\frac{\psi_{\downarrow}^{-}(\mathbf{k}) \psi_{\downarrow}^{+}(-\mathbf{k})}{\omega+i0 \sign{\omega} -\epsilon_{45,\pm}}\right] \nonumber\\
&=&  \frac{ig}{4\sqrt{\pi}l_B^2} \int \frac{d\omega\, dk_z}{(2\pi)^3} \Bigg\{\frac{1}{\omega^2+i0 -\left[\mu+ v_F (k_z -b_z)\right]^2-|\Delta_{x}|^2}  +\frac{1}{\omega^2+i0 -\left[\mu+ v_F (k_z +b_z)\right]^2-|\Delta_{x}|^2} \Bigg\}.
\end{eqnarray}
Let us concentrate on the first term in the curly brackets. The contribution from the second term is the same albeit with $b_z\to-b_z$. We have
\begin{eqnarray}
\label{App-B-gap-equation-1-first-v1}
&&\int d\omega\, \int_{-\Lambda}^{\Lambda} dk_z  \frac{1}{\omega^2+i0 -\left[\mu+ v_F (k_z -b_z)\right]^2-|\Delta_{x}|^2} =-i \pi \int d\omega\, \int_{-\Lambda}^{\Lambda} dk_z  \delta \left[\omega^2 -\left[\mu+ v_F (k_z -b_z)\right]^2-|\Delta_{x}|^2\right]\nonumber\\
&&=-i \pi \int_{-\Lambda}^{\Lambda} dk_z  \frac{1}{\sqrt{\left[\mu+ v_F (k_z -b_z)\right]^2+|\Delta_{x}|^2}} =-\frac{i \pi}{v_F} \ln{\left|\frac{v_F\Lambda +(\mu - v_F b_z) +\sqrt{\left(v_F\Lambda +\mu-v_F b_z\right)^2+|\Delta_{x}|^2}}{-v_F\Lambda +(\mu - v_F b_z) +\sqrt{\left(v_F\Lambda -\mu+ v_F b_z\right)^2+|\Delta_{x}|^2}} \right|}
\nonumber\\
&&\approx -\frac{i \pi}{v_F} \ln{\left(\frac{4v_F^2\Lambda^2}{|\Delta_{x}|^2}\right)},
\end{eqnarray}
where $\Lambda$ is the momentum cutoff. The gap equation (\ref{App-B-gap-equation-1}) reads
\begin{eqnarray}
\label{App-B-gap-equation-2-v1}
\Delta_x\approx \frac{g\Delta_x}{2\sqrt{\pi}(2\pi)^2 l_{B}^2 v_F}  \ln{\left(\frac{4v_F^2\Lambda^2}{|\Delta_{x}|^2}\right)}.
\end{eqnarray}
Its nontrivial solution is
\begin{eqnarray}
\label{App-B-gap-equation-sol-v1}
|\Delta_{x}| = 2 v_F\Lambda\, \mbox{exp}\left[-\frac{\sqrt{\pi} (2\pi)^2 l_B^2 v_F}{g}\right].
\end{eqnarray}
This result is similar to the dynamical gap generation in the framework of the magnetic catalysis~\cite{Gusynin:1994,Miransky-Shovkovy:rev-2015}.

In summary, the energy spectrum and the gap generation in a superconducting Weyl semimetal in a strong magnetic field $B$ are qualitatively different from the case of a strong pseudomagnetic field $B_5$ discussed in Sec.~\ref{sec:pairing}. Indeed, while no true energy gap is opened for an inter-node s-wave pairing in a strong pseudomagnetic field, a nontrivial solution to the gap equation exists at $B\neq0$ and a true gap in the energy dispersion is realized (cf. Figs.~\ref{fig:energy-dispersion-B5} and \ref{fig:energy-dispersion-B}). We would like to note, however, that the above analysis lacks a self-consistent treatment of electromagnetism and, consequently, does not take into account the Meissner effect. Therefore, it is included only for comparison of the superconducting pairings in the cases $B_5\neq0$ and $B\neq0$.

\section{Shorthand notation for spectral and Green's functions}
\label{sec:App-1}

In this appendix, we present a shorthand notation for the spectral function components used in Sec.~\ref{sec:tunneling-current} in the main text. In particular, we used the following notation in Eq.~(\ref{tc-IV-expl-MB}):
\begin{eqnarray}
\label{tc-shorthand-A}
A_{W/S}(\omega, \mathbf{k}) &=& -\frac{1}{2\pi} \left[G_{W/S}^{\rm R}(\omega, \mathbf{k})-G_{W/S}^{\rm A}(\omega, \mathbf{k})\right]_{\mu=0},\\
\label{tc-shorthand-C}
C_{W/S}(\omega, \mathbf{k}) &=& -\frac{1}{2\pi} \left[F_{W/S}^{\rm R}(\omega, \mathbf{k})-F_{W/S}^{\rm A}(\omega, \mathbf{k})\right]_{\mu=0},\\
\label{tc-shorthand-Cdag}
C_{W/S}^{\dag}(\omega, \mathbf{k}) &=& -\left[C_{W/S}(\omega, \mathbf{k})\right]^{\dag},\\
\label{tc-shorthand-Abar}
\bar{A}_{W/S}(\omega, \mathbf{k}) &=& -\frac{1}{2\pi} \left[\bar{G}_{W/S}^{\rm R}(\omega, \mathbf{k})-\bar{G}_{W/S}^{\rm A}(\omega, \mathbf{k})\right]_{\mu=0}.
\end{eqnarray}
Here the normal and anomalous Green's functions for quasiparticles in a superconductor defined by Hamiltonian (\ref{PSE-2-HS-def}) are
\begin{eqnarray}
\label{tc-shorthand-SC-G}
G_{S}^{R}(\omega,\mathbf{k}) &=& \frac{\omega+\xi_{\mathbf{k}}}{\omega^2+i0\sign{\omega}-\xi_{\mathbf{k}}^2 -|\Delta|^2},\\
\label{tc-shorthand-SC-F}
F_{S}^{R}(\omega,\mathbf{k}) &=& \frac{\Delta}{\omega^2+i0\sign{\omega}-\xi_{\mathbf{k}}^2 -|\Delta|^2},\\
\label{tc-shorthand-SC-Gbar}
\bar{G}_{S}^{R}(\omega,\mathbf{k}) &=& \frac{\omega-\xi_{\mathbf{k}}}{\omega^2+i0\sign{\omega}-\xi_{\mathbf{k}}^2 -|\Delta|^2}.
\end{eqnarray}
In a Weyl semimetal, we have (see also Sec.~\ref{sec:pairing})
\begin{eqnarray}
\label{tc-shorthand-W-G-28}
G_{W,28}^{R}(\omega,k_y, k_z,x,x) &=& \sum_{\pm}\frac{|N_{+}|^2e^{-\frac{1}{l_{B_5}^2}\left[l_{B_5}^2 (k_y -b_y) +x\right]^2}}{\omega +i0 -\epsilon_{28,\pm}},\\
\label{tc-shorthand-W-G-35}
G_{W,35}^{R}(\omega,k_y, k_z,x,x) &=& \sum_{\pm}\frac{|N_{-}|^2e^{-\frac{1}{l_{B_5}^2}\left[l_{B_5}^2 (k_y +b_y) +x\right]^2}}{\omega +i0 -\epsilon_{35,\pm}},\\
\label{tc-shorthand-W-Gbar-28}
\bar{G}_{W,28}^{R}(\omega,k_y, k_z,x,x) &=& \sum_{\pm}\frac{|N_{+}|^2e^{-\frac{1}{l_{B_5}^2}\left[-l_{B_5}^2 (k_y +b_y) +x\right]^2}}{\omega +i0 -\epsilon_{28,\pm}},\\
\label{tc-shorthand-W-Gbar-35}
\bar{G}_{W,35}^{R}(\omega,k_y, k_z,x,x) &=& \sum_{\pm}\frac{|N_{-}|^2e^{-\frac{1}{l_{B_5}^2}\left[-l_{B_5}^2 (k_y -b_y) +x\right]^2}}{\omega +i0 -\epsilon_{35,\pm}},
\end{eqnarray}
\begin{eqnarray}
\label{tc-shorthand-W-F-28}
F_{W,28}^{R}(\omega,k_y, k_z,x,x) &=& \sum_{\pm}\frac{|N_{+}|^2e^{-\frac{1}{l_{B_5}^2}\left[-l_{B_5}^2 (k_y +b_y) +x\right]^2}}{\omega +i0 -\epsilon_{28,\pm}} \frac{\Delta_z}{\mu -\epsilon_{28,\pm} -v_F (k_z-b_z)},\\
\label{tc-shorthand-W-F-35}
F_{W,35}^{R}(\omega,k_y, k_z,x,x) &=& \sum_{\pm}\frac{|N_{-}|^2e^{-\frac{1}{l_{B_5}^2}\left[-l_{B_5}^2 (k_y -b_y) +x\right]^2}}{\omega +i0 -\epsilon_{35,\pm}} \frac{\Delta_z}{\mu -\epsilon_{35,\pm} -v_F (k_z+b_z)}.
\end{eqnarray}
Here $N_{+}$ is the normalization factor defined in Eq.~(\ref{wave-functions-norm-28-1}), $N_{-}$ is given by the same expression albeit with $\mathbf{b}\to-\mathbf{b}$, and $l_{B_5}=\sqrt{c/|eB_5|}$ is the pseudomagnetic length.

\end{document}